%% file: BKGjelsten_ATLAS_IDM2008_proceedings.tex
\documentclass{PoS}

\title{R-parity conserving SUSY searches in ATLAS (25'+5')}

\ShortTitle{R-parity conserving SUSY searches in ATLAS (25'+5')}

\author{
\speaker{B\o rge Kile GJELSTEN}%
         \thanks{On behalf of the ATLAS collaboration.}\\
        University of Oslo\\
        E-mail: \email{B.K.Gjelsten@fys.uio.no}}

\abstract{Searches for physics beyond the Standard Model of particle physics are about to enter a new era with the startup of the Large Hadron Collider (LHC) at CERN. 
Prospects for R-parity conserving supersymmetry discovery and measurements with the ATLAS detector at the LHC are presented. 
Methods for determining the parameters of the underlying supersymmetry model and the relation to dark matter estimation are discussed.}

\FullConference{Identification of dark matter 2008\\
		 August 18-22, 2008\\
		 Stockholm, Sweden}

\input{bdefs.tex}

\def\ifb{fb$^{-1}$}
\def\EtMiss{E_{T}^{\rm miss}}
\def\Meff{M_{\rm eff}}
\def\mTTwo{M_{T2}}
\def\mtautau{m_{\tau\tau}}
\def\mqR{m_{\qR}}

\def\sle{\tilde{\ell}}
\def\msle{m_{\sle}}
\def\mll{m_{\ell\ell}}
\def\mllq{m_{\ell\ell q}}
\def\mlq{m_{\ell q}}

\usepackage[abs]{overpic}
\usepackage{graphicx}

\begin{document}

\section{Introduction}

A long standing problem of fundamental physics 
regards the status and features of dark matter, 
a kind of matter not accounted for in the Standard Model (SM) 
of particle physics 
but needed to explain various cases in astronomy and cosmology 
like the observed rotation of galaxies. 
In later years dark matter predictions 
have been brought to very high precision 
with the measurement of anisotropies 
in the cosmic microwave background \cite{WMAPall}.

One of the most promising candidates for a theory beyond the Standard Model 
which also provides a viable candidate for dark matter 
is R-parity conserving supersymmetry (SUSY). 
In these models the spectrum of elementary particles is extended 
to include 
supersymmetric partners which differ from their Standard Model counterparts 
by half a unit of spin. 
Accompanying the quarks, leptons, gluons, gauge bosons and higgses 
(of the extended Higgs sector) 
there will be 
squarks, sleptons, gluinos, gauginos and higgsinos, respectively. 
The two latter types have overlapping quantum numbers 
and mix to give charginos and neutralinos mass eigenstates.

To comply with experimental non-observation 
(no spin-0 electron has ever been observed) 
the sparticles must be heavier than the Standard Model particles: 
supersymmetry must be broken. 
Our ignorance of the origin of supersymmetry breaking introduces 
$>$100 free parameters (masses, mixing angles, phases) 
into the low-energy Lagrangian. 
Some of the new couplings give rapid proton decay 
unless considerably constrained. 
The addition of a new, conserved symmetry, R-parity, 
removes all such dangerous terms. 
The introduction of R-parity has the important side effect that 
any interaction must involve an even number of superpartners. 
This means that there will be a Lightest Supersymmetric Particle (LSP) 
which is stable and hence a candidate for dark matter. 
In most relevant cases, and assumed in this paper, 
the lightest neutralino ($\NO$) will be the LSP.

The Minimal Supersymmetric Model (MSSM), which is the minimal implementation 
of supersymmetry onto the Standard Model 
with no additional assumptions except R-parity conservation, 
contains 105 
undetermined parameters. 
To avoid Flavour-Changing Neutral Currents (FCNC) and 
CP violation beyond experimental values, 
assumptions on the many free parameters are usually made, 
effectively reducing the dimension of parameter space to handleable proportions,
and also rendering the models more apt for phenomenological studies. 
So-called Minimal Supergravity (mSUGRA), 
contains only four continuous parameters, 
all given at the GUT scale: 
$\mZero$ and $\mHalf$, 
the masses of the spin-0 and the spin-1/2 
sparticles, respectively, 
$\AZero$, the trilinear coupling, 
$\tanB$, the vacuum-expectation value (VEV) of the neutral component of the Higgs doublet giving masses to up-type particles divided by the VEV of the neutral component of the Higgs doublet giving masses to down-type particles, 
as well as the sign of the Higgs potential parameter $\mu$. 
%
The investigations reported on in this write-up 
are conducted within mSUGRA unless otherwise specified.

Despite decades of searches at the major experimental sites 
there is yet no evidence that nature is supersymmetric. 
Instead limits are set, mainly by the non-observation at LEP and Tevatron. 
%
%
%
With the startup of the Large Hadron Collider (LHC) 
a new energy regime will be explored. 
While the Tevatron has $p\bar p$ collisions at 2~TeV center of mass energies, 
the LHC is constructed to collide protons on protons at 14~TeV. 
The relevant center of mass energy is however that of the 
colliding quarks/antiquarks/gluons, constituents of the proton, 
yielding an effective center of mass energy of the order of a few TeV.  
Nevertheless, 
if supersymmetry is to address the hierarchy problem, 
as is one of its main motivations, 
supersymmetric particles should have masses not very much above 1~TeV, 
and hence in most cases be accessible by the effective 
center-of-mass energies of the LHC.

\section{Inclusive searches}

With R-parity conserved, 
sparticles will be produced in pairs at colliders. 
If not too heavy, squarks and gluinos will dominate the SUSY production 
at LHC due to the strong interactions. 
Each of the initial sparticles will then set off a cascade of decays into 
lighter and lighter sparticles, 
at each step splitting off Standard Model particles;  
gluons, quarks and leptons as well as W's, Z's, Higgses and photons. 
The LSP, neutral and weakly interacting, 
will leave the detector without a trace, 
but not traceless. 
While each hard-collision in a hadron-hadron machine will have a 
varying, non-zero boost in the beam direction 
leaving this variable insensitive to the non-detected LSPs, 
the momenta in the transverse direction should sum up to zero. 
The transverse energy carried away by the two escaping LSPs 
produces the most generic signature of R-parity conserving SUSY:  
missing transverse energy ($\EtMiss$). 
Additional signatures are several hard jets and possibly leptons, 
both produced directly in the cascade or in the decay of intermediate 
W, Z or Higgs bosons. 
(Hard photons are crucial in SUSY models where the breaking is 
mediated by gauge interactions, not considered in this paper.) 
Natural SUSY search channels are therefore based on 
considerable $\EtMiss$, $N$ hard jets and $M$ leptons. 
A simple variable incorporating all of these features 
and therefore apt to distinguish SUSY from SM events 
is the effective mass ($\Meff$), 
defined as the scalar sum of $\EtMiss$ and the transverse 
momentum of each of the jets and leptons.

\begin{figure}[htb]

\hspace{1em}
\begin{overpic}[height=0.285\textheight,width=0.45\textwidth]{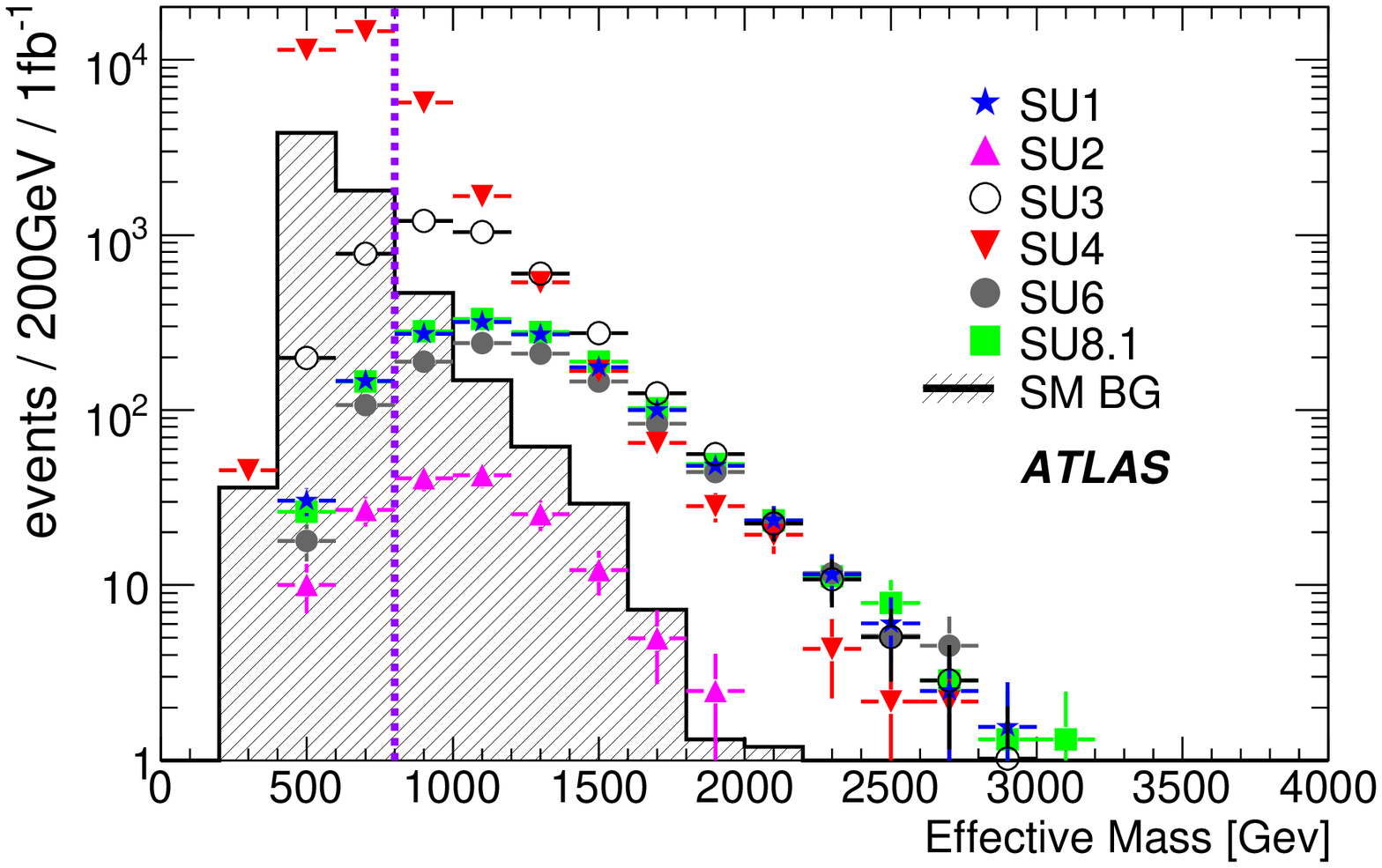}
\put(210,0){\includegraphics[width=0.5\textwidth]{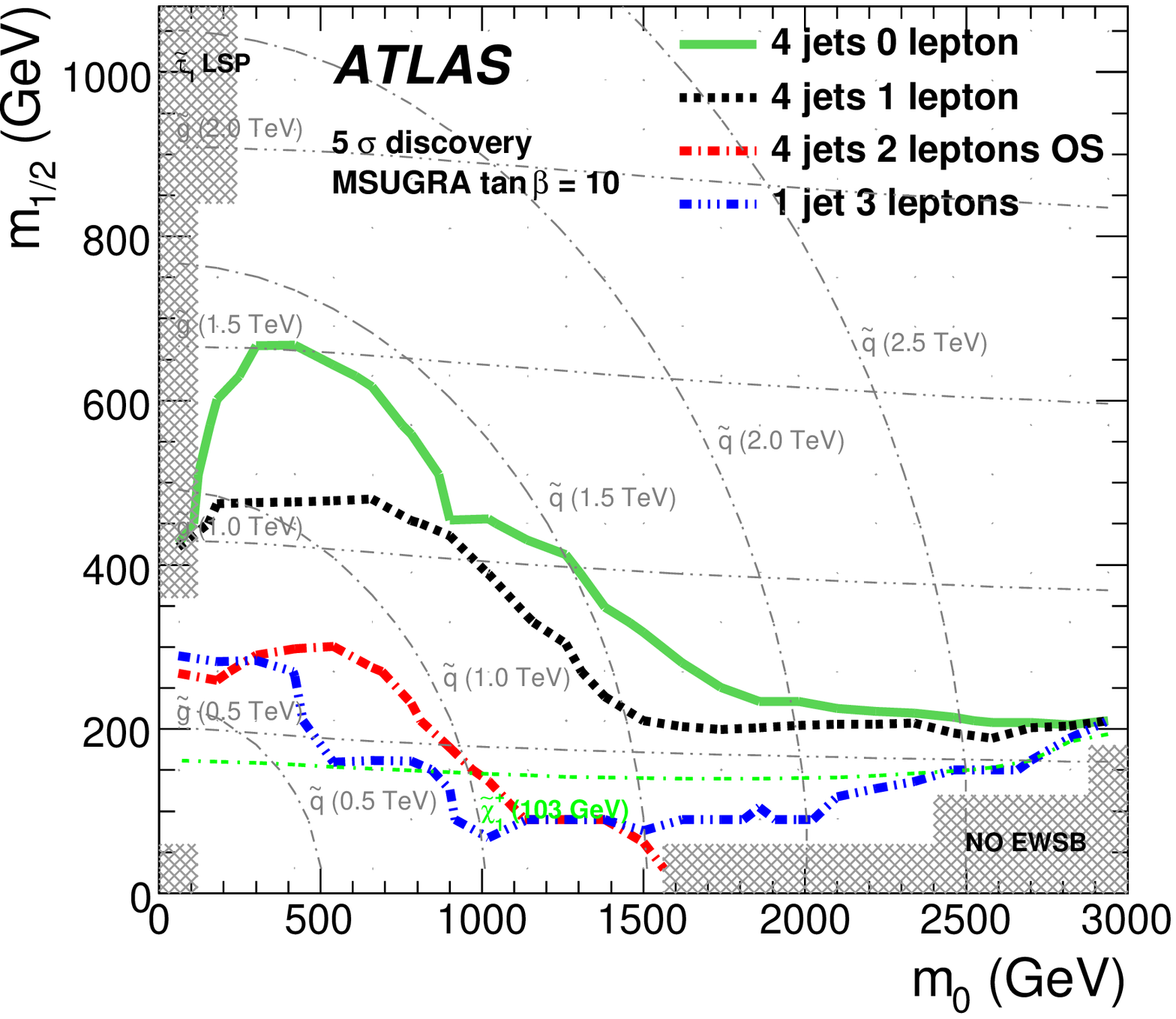}}
\put(114,72){\small PRELIMINARY} 
\put(320,72){\small PRELIMINARY} 
\end{overpic}

\caption{Left: $\Meff$ distribution for mSUGRA benchmark scenarios SUx and SM in the 0-lepton 4-jet channel. 
See \cite{ATLASBOOK} for definition of the SUx scenarios. 
SUSY signal is visible over SM for most scenarios. 
Right: 5-sigma mSUGRA discovery contours. 
Both plots correspond to 1~\ifb\ of integrated luminosity.
}
\label{fig:0L4J}
\end{figure}

Figure~\ref{fig:0L4J} (left) shows $\Meff$ for a set of mSUGRA benchmark points 
in the 0-lepton 4-jet channel. 
Most of the benchmark scenarios show a noticeable excess of events at high 
$\Meff$ values already with 1~\ifb, which is of the order of the integrated 
luminosity expected to be collected in 2009--2010. 
The right plot shows expected 5-sigma discovery contours 
with 1~\ifb of ATLAS data 
for mSUGRA in $\mZero$ and $\mHalf$ with 
$\tanB=10$, $\AZero=0$ and $\mu>0$ fixed. 
Selection cuts are detailed in \cite{ATLASBOOK}, 
where also many other search channels are 
discussed. 
If nature is super\-symmetric with a reali\-sation 
giving pheno\-menologies not too different from mSUGRA 
(in particular in $\EtMiss$ and jet hardness), 
and the SUSY scale is such that the hierarchy problem is addressed, 
then ATLAS is likely to make discoveries 
already with 1~\ifb\ which might be reached in the first years of operation.

\section{SUSY dark matter}

To confirm that an excess above the SM, 
as could be observed e.g.\ in the $\Meff$ distribution, 
Figure~\ref{fig:0L4J} (left), 
is in fact due to SUSY, 
exclusive studies would be needed 
in which particular decay chains are sought isolated. 
This is also what is needed in order to estimate the supersymmetric 
contribution to dark matter. 
In the very early universe the LSPs would be in thermal equilibrium, 
$\NO\NO\rightleftharpoons f\bar f$. 
As the universe expanded and the temperature eventually sank 
below the LSP mass, 
annihilation of LSPs would start to dominate over their production. 
Then, at some stage, with the steady reduction of the LSP density, 
the collision rate would effectively come to a stop.  
This is the moment of LSP freeze-out. 
The amount of LSP relic density of the universe today 
is determined by the efficiency of the LSP annihilation 
before freeze-out, 
i.e.\ by the cross section for 
$\NO\NO\to f\bar f/W^+W^-/ZZ$ 
(and also for co-annihilation, e.g.\ $\NO\stau\to\tau\gamma$). 

In order to determine the LSP relic density from ATLAS data, 
it is therefore critical to estimate the SUSY parameters which 
enter the dominating annihilation process(es). 
In different parts of SUSY parameter space, 
different annihilation processes are dominating. 
In mSUGRA the LSP (the lightest neutralino, $\NO$) 
is usually mostly bino, i.e.\ the partner of the U(1) gauge boson. 
In this case the basic annihilation process $\NO\NO\to f\bar f$ 
through the exchange of a slepton (or squark) 
usually has too low cross section. 
Unless the LSP and slepton (squark) masses are sufficiently low (bulk region), 
this annihilation process 
will not alone be enough to reduce the LSP density 
to the required amounts 
before freeze-out, 
resulting in too large relic density, 
already excluded by current dark matter estimates. 
This situation also ports into non-mSUGRA models. 
One way to increase the annihilation rate, 
is to have the bino-content of the LSP 
diluted by wino (SU(2) coupling) and/or higgsino components. 
In mSUGRA scenarios only the latter possibility is allowed 
(focus point region). 
Outside of mSUGRA 
both possibilities can occur with ease. 
Another way in which the annihilation cross section will 
be enhanced is through resonant s-channel scattering with an intermediate 
Higgs boson. 
For this to happen $m_{h/H/A}\approx 2 m_{\NO}$ must be satisfied 
(Higgs funnel region). 
Yet another way is if the next-to-lightest sparticle(s) (NLSP) 
is not much heavier than the LSP. Then co-annihilation of 
LSP+NLSP assists in reducing the LSP density to the required amounts.

Some of the most relevant quantities to measure (or control) are therefore the 
mass of the LSP, 
of the (lightest) sleptons/squarks 
(and whether they are superpartners of left-handed or right-handed SM particles), 
the neutral Higgs masses. 
Furthermore, to determine the LSP composition 
the mass of additional neutralinos are important. 

\section{Endpoints}
Reconstructing SUSY events 
from the measured quantities, jets, leptons and $\EtMiss$ 
is complicated by the undetected LSPs 
(and the measured $\EtMiss$ 
cannot easily be split between them) 
as well as there being two decay chains. 
The latter adds combinatorial background, 
the former requires the use of endpoints rather than mass peaks 
to determine masses. 

In large parts of SUSY parameter space 
the next-to-lightest neutralino ($\NT$) 
is lighter than 
the initially produced squarks (and gluinos). 
This allows for the golden decay chain 
$\sq \to{q\NT} {(\to q\ell\sle)} \to q\ell\ell\NO$
which is particularly useful when $\ell$ denotes electrons or muons, 
since these are very accurately reconstructed in ATLAS. 
(The parenthesis is opened if $\mNT>\msle$. If $\mgl>\msq$, 
an additional jet may come into useful play from 
an initial $\gl\to q\sq$.) 
With such a chain isolated, four invariant masses can be 
constructed from the three available four-vectors, 
$\mll$, $\mllq$ and two copies of $\mlq$ (which can be kept apart). 
Add more events, and four distributions are obtained. 
While the peaks of these distribution have no significant interpretation, 
the endpoints do, and are given by simple formulas 
involving the masses of the intermediate sparticles. 
%
%
With sufficient endpoints measured the relevant sparticle masses 
can be determined.

Similar techniques can be used on the decay $\gl\to t\tO\to tb\CO$ 
to constrain the mass of the lighter stop, which if sufficiently light 
would add an important co-annihilation processes. 
The stau mass(es) can be constrained by 
$\NT\to\tau\stau\to\tau\tau\NO$ if open, 
although with precision much inferior to the selectron/smuon case. 
For sleptons with mass above $\NT$, mass constraints might be obtained from 
an $\mTTwo$ 
(see \cite{ATLASBOOK}) 
analysis of direct slepton production $\sle\sle\to\ell\ell\NO\NO$. 
Similar techniques may be used to obtain the mass of the right-handed squark ($\qR$). 
Comparison of branching ratios is another instrument to obtain 
information on quantities of relevance to LSP relic density estimates. 

Some of these measurements will need many years of LHC operation, 
others can be made the first year, 
although with limited precision.

\section{``First year'' luminosity, 1.0~\ifb}

For the mSUGRA benchmark point SU3 the golden decay chain 
will give important information already at 1~\ifb. 
Figure~\ref{fig:endpointsSU3} shows two of the available mass distributions. 

\begin{figure}[htb]
\hspace{1em}
\begin{overpic}[width=0.45\textwidth]{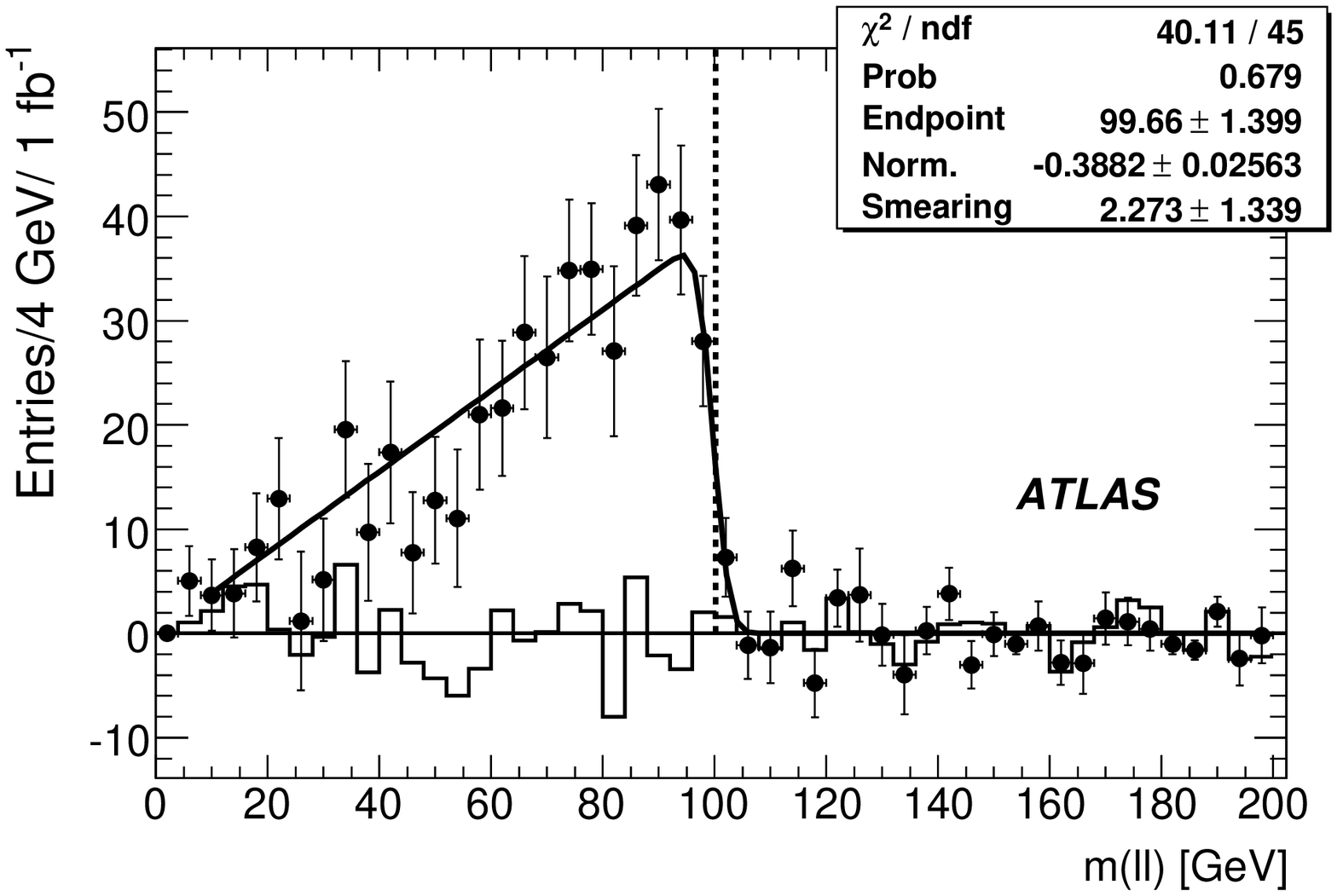}
\put(210,0){\includegraphics[width=0.45\textwidth]{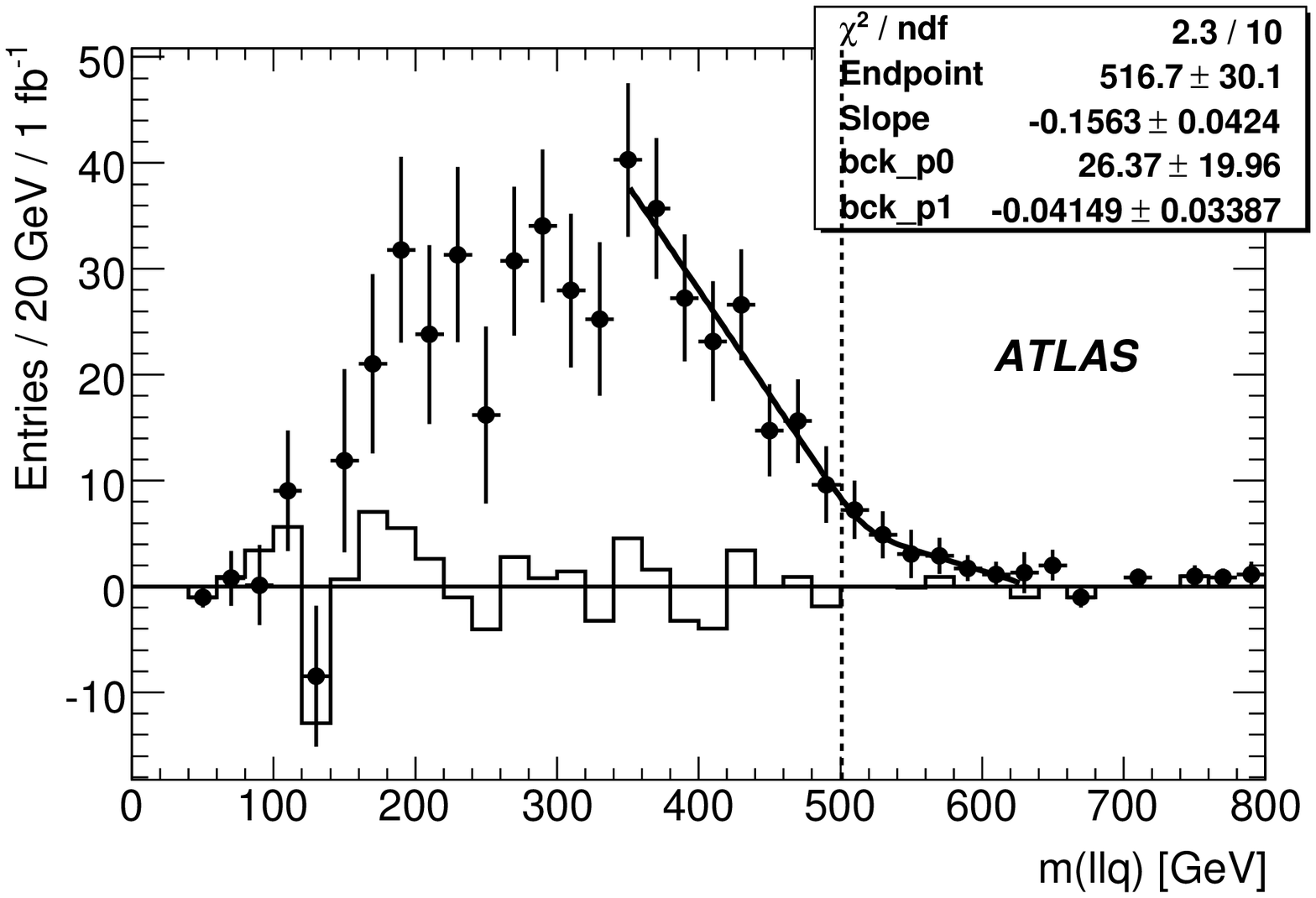}}
\put(110,86){\small PRELIMINARY} 
\put(322,86){\small PRELIMINARY} 
\end{overpic}
\caption{Distributions of $\mll$ and $\mllq$ for benchmark point SU3 at 1~\ifb\ integrated luminosity. 
(Bins of negative value occur from a subtraction technique where events of opposite-flavour are subtracted from the same-flavour ones, which effectively removes large parts of the background.) 
The vertical lines mark the theoretical endpoints. The distributions fall off close to the theoretical endpoint. 
}
\label{fig:endpointsSU3}
\end{figure}

The edges are clear and can be fitted with appropriate functions, 
although more luminosity will clearly improve the precision. 
The two $\mlq$ distributions as well as an additional $\mllq$ distribution 
also give endpoint measurements, allowing the masses to be determined 
although at quite low precision. 
Some other SU3 measurements which can be made at 1~\ifb\ 
are the $\mtautau$ endpoint, 
the mass of the right-handed squark 
as well as an endpoint involving the lighter stop, 
see \cite{ATLASBOOK} for details.

Combining the endpoints of the golden chain 
with the $\mqR$ measurement, an mSUGRA scan can be performed 
to constrain the SUSY parameters. 
Table~\ref{tab:scanSU3} shows the result. 
The mass parameters $\mZero$ and $\mHalf$ are already accurately determined, 
with 9\% and 2\% uncertainty, respectively, 
and the sign of $\mu$ is correctly found. 
The two other parameters, $\tanB$ and in particular $\AZero$ 
are however not very well constrained by the 1~\ifb\ measurements. 

\begin{table}[htb]
\caption{Nominal and estimated mSUGRA parameters for SU3 at 1~\ifb.}
\label{tab:scanSU3}
  \begin{center}
    \begin{tabular}{lccc}
      \hline
      Parameter & SU3 value & fitted value & exp. unc. \\
      \hline
      $\tan\beta$ & $6$       & 7.4         & 4.6           \\
      $m_{0}$     & $\phantom{-}100$ GeV & \phantom{1}$98.5$ GeV & $\pm 9.3$ GeV \\
      $m_{1/2}$   & $\phantom{-}300$ GeV & $317.7$ GeV & $\pm 6.9$ GeV \\
      $A_{0}$     &$-300$ GeV & \phantom{1}$445$ GeV   & $\pm 408$ GeV \\
      \hline
    \end{tabular}
  \end{center}
\end{table}

The final step, estimation of the relic density, 
is not made in this study. 
While it seems plausible that the relic density 
under the assumption of mSUGRA  
could be determined with sufficient precision that a comparison 
with WMAP values would make sense, 
going outside of mSUGRA probably will require considerably 
higher integrated luminosities.

\section{``Ultimate'' luminosity, 300~\ifb}

In \cite{DM_PT} the LSP relic density of the mSUGRA benchmark point SPS~1a, 
quite similar to 
the SU3 point studied above, 
is estimated from ATLAS measurements after an ``ultimate'' 
integrated luminosity of 300~\ifb. 
In this study an mSUGRA scenario is assumed from the start, 
and a top-down approach is taken, 
i.e.\ simply scanning over the SUSY parameters. 
With the high precision on the ATLAS measurements after 300~\ifb, 
the mSUGRA parameters are well determined, 
$
\sigma(\mZero)/\mZero = 2\%,\ 
\sigma(\mHalf)/\mHalf = 0.6\%,\ 
\sigma(\tanB) / \tanB = 9\%,\ 
\sigma(\AZero)/\AZero = 16\%,\
$
yielding a precision on the LSP relic density of $\sim$3\%. 

While such a scan will undoubtedly be performed should ATLAS 
provide the relevant measurements, 
it would be unwise to rely too much on nature having settled on 
such a well-behaving model. 
Non-consistency with mSUGRA assumptions would be revealed 
by bad likelihood values of the parameter fit, 
provided the measurements come with sufficient precision. 
Less constrained models would then need to be considered.

In \cite{DM_NPT} the same objective is undertaken for a 
similar benchmark point, SPS~1a', 
and the same integrated luminosity, but with the 
mSUGRA assumptions relaxed to 24-parameter MSSM. 
A bottom-up approach is attempted 
in which each sparticle sector is reconstructed as much as the measurements allow, 
then the remaining degrees of freedom are scanned over to monitor the 
sensitivity of the relic density to the undetermined parameters. 
The neutralino sector is attacked first. 
The masses of $\NO$ and $\NT$, determined from the endpoints, 
provide two constraints on the four parameters defining the sector. 
For SPS~1a' also the mass of $\NFour$ can be determined, 
leaving only one parameter unconstrained, here taken as $\tanB$. 
Next, the slepton sector is attacked, 
which turns out somewhat more involved. 
Out of several unsettled issues 
the largest effect on the relic density is however found to 
come from the uncertainty with which the mass of the lightest stau 
can be determined (i.e.\ the uncertainty on the endpoint $\mtautau$). 
Finally the Higgs sector is investigated, mainly 
to assess the possibility of LSP annihilation through Higgs resonance. 
The final precision on the relic density is found to be 20\%(10\%) 
for $\sigma(\mtautau)=5(1)$~GeV. 
The results are less precise than the mSUGRA precision of 3\%, 
but nevertheless impressive. 
A top-down analysis for 24-parameter MSSM on the same SPS~1a' measurements 
confirms the results\cite{DM_Peskin}.

\section{Summary}

If supersymmetry is to have a word on the hierarchy problem, 
and is realised with phenomenologies 
(in particular in the amount of $\EtMiss$ and jet hardness) 
comparable to the mSUGRA models widely studied, 
ATLAS has good opportunities of making discoveries 
in the early phases of operation. 
For favourable scenarios measurements 
of exclusive quantities 
can be made with moderate precision already in the first year, 
and with very high precision after some years. 
For less favourable scenarios more luminosity is needed to start 
characterising the specific instance of supersymmetry. 
The accuracy with which the LSP relic density can be estimated, 
depends quite a lot on 
our location in SUSY parameter space. 
With ultimate luminosity an accuracy of 10\% can be achieved 
in very favourable cases. 
In other cases very little can be claimed. 
There is however an interesting asymmetry in the problem. 
If, from ATLAS measurements, one of the annihilation processes 
is found 
to reduce the relic density below the measured dark matter density, 
we would already know, 
regardless of the status of the other annihilation processes, 
that SUSY alone does not account for all the dark matter. 
Also, if one of the processes is found to give LSP relic density in 
good agreement with the current dark matter limits 
(with the cross section of the other processes remaining unknown), 
temptation will be large 
to discard the possibility of this being accidental, 
and hence suggest that the (principal) dark matter content of the universe 
is indeed supersymmetric.  
A final word on supersymmetry at colliders and dark matter: 
stability of the LSP at ATLAS time scales, 
as testified by large amounts of $\EtMiss$ does not guarantee 
LSP stability at the time scale of the universe.


\end{document}


%% file: bdefs.tex
\def\BI{\begin{itemize}}
\def\EI{\end{itemize}}
\def\BE{\begin{enumerate}}
\def\EE{\end{enumerate}}
\def\BA{\begin{array}}
\def\EA{\end{array}}
\def\BF{\begin{figure}}
\def\EF{\end{figure}}
\def\BC{\begin{center}}
\def\EC{\end{center}}
\def\BP{\begin{picture}}
\def\EP{\end{picture}}

\def\NO{\tilde\chi^0_1} 
\def\NT{\tilde\chi^0_2} 

\def\NFour{\tilde\chi^0_4}

\def\gl{\tilde g}
\def\sq{\tilde q}

\def\stau{\tilde\tau}

\def\CO{\tilde\chi_1^\pm}

\def\qR{\tilde q_R}

\def\sle{\tilde l}

\def\tO{\tilde t_1}

\def\mNT{m_{\NT}}

\def\mgl{m_{\gl}}

\def\msle{m_{\sle}}

\def\msq{m_{\sq}}

\def\mHalf{m_{1/2}}
\def\mZero{m_0}
\def\AZero{A_0}

\def\mll{m_{ll}}

\def\tanB{\tan\beta}


%% file: BKGjelsten_ATLAS_IDM2008_proceedings.bbl
\begin{thebibliography}{99}
  
\bibitem{WMAPall} Papers of the five-year WMAP observations are available at: 
http://lambda.gafc.nasa.gov/product/map/current/map\_bibliography.cfm.

\bibitem{ATLASBOOK} ATLAS Collaboration, {\it Expected Performance of the ATLAS Experiment, Detector, Trigger and Physics}, CERN-OPEN-2008-020, Geneva, 2008, to appear.
  
\bibitem{DM_PT} G.~Polesello and D.~R.~Tovey, 
{\em Constraining SUSY Dark Matter with the ATLAS Detector at the LHC}, 
JHEP 0405 (2004) 071 [hep-ph/0403047]

\bibitem{DM_NPT} M.~M.~Nojiri, G.~Polesello and D.~R.~Tovey, 
{\em Constraining Dark Matter in the MSSM at the LHC}, 
JHEP 0603 (2006) 063 [hep-ph/0512204]

\bibitem{DM_Peskin} E.~A.~Baltz, M.~Battaglia, M.~E.~Peskin and T.~Wizansky, 
{\em Determination of Dark Matter Properties at High-Energy Colliders}, 
Phys. Rev. D 74, 103521 (2006) [hep-ph/0602187]


\end{thebibliography}
